\documentstyle[]{mn}

\newif\ifAMStwofonts
\AMStwofontstrue

\def\etal{{\it et al.}}

\def\eg{{\it e.g.}}
\def\lap{\hbox{${_{\displaystyle<}\atop^{\displaystyle\sim}}$}}
\def\gap{\hbox{${_{\displaystyle>}\atop^{\displaystyle\sim}}$}}

\newcommand{\nd}        {\mbox{\boldmath$n_{d}$}}
\newcommand{\J}         {\mbox{\boldmath$J$}}

\newcommand{\bfI}       {\mbox{\boldmath $I$}}
\newcommand{\bfdelta}   {\mbox{\boldmath $\delta$}}
\newcommand{\Om}        {\mbox{\boldmath $\Omega$}}
\newcommand{\nom}       {\mbox{\boldmath$n_{\Omega}$}}

\newcommand{\nJ}        {\mbox{\boldmath$n_{J}$}}
\newcommand{\nx}        {\mbox{\boldmath$n_x$}}
\newcommand{\ny}        {\mbox{\boldmath$n_y$}}
\newcommand{\nz}        {\mbox{\boldmath$n_z$}}
\newcommand{\Oms}       {\mbox{\boldmath$\Omega_s$}}
\newcommand{\bfu}         {\mbox{\boldmath$u$}}
\newcommand{\fbf}         {\mbox{\boldmath$f$}}
\newcommand{\rbf}         {\mbox{\boldmath$r$}}

\newcommand{\Nbf}         {\mbox{\boldmath$N$}}

\title{Vortex Unpinning in Precessing Neutron Stars} 
\author[B.~Link and C.~Cutler]
{B.~Link$^1$ and C.~Cutler$^2$ \\
$^1$Montana State University, Department of Physics, Bozeman MT
59717; blink@dante.physics.montana.edu \\
$^2$Max-Planck-Institut fuer Gravitationsphysik, Golm bei Potsdam,
Germany; cutler@aei-potsdam.mpg.de}

\pagerange{1--6}
\pubyear{2001}

\begin{document}

\topmargin = -2pc
\maketitle

\label{firstpage}

\begin{abstract}
The neutron vortices thought to exist in the inner crust of a neutron
star interact with nuclei and are expected to pin to the nuclear
lattice. Evidence for long-period precession in pulsars, however,
requires that pinning be negligible. We estimate the strength of
vortex pinning and show that hydrodynamic forces present in a
precessing star are likely sufficient to unpin all of the vortices of
the inner crust. In the absence of precession, however, vortices could
pin to the lattice with sufficient strength to explain the giant
glitches observed in many radio pulsars.
\end{abstract}


\section{Introduction}

Observations of precession of neutron stars probe the manner in which
the stellar crust is coupled to the liquid interior.  Stairs, Lyne \&
Shemar (2000) recently presented strong evidence for precession of an
isolated neutron star at a period of $\simeq 1000$ d, 500 d or 250 d,
with correlated variations in the pulse duration. The wobble angle of
the precession is inferred to be $\simeq 3^\circ$ (Link \& Epstein
2001). Less compelling evidence for precession is seen from other
neutron stars. Quasi-periodic timing residuals are exhibited by PSR
B1642-03 (Shabanova, Lyne \& Urama 2001), in association with roughly
cyclical changes in pulse duration (Cordes 1993).  Quasi-periodic
timing residuals are seen in the Crab (Lyne, Pritchard \& Smith 1988;
\v{C}ade\v{z} \etal\ 2001) and Vela pulsars (Deshpande \& McCulloch
1996), though associated changes in pulse duration are not seen. The
35-d periodicity seen in the accreting system Her X-1 (Tannanbaum
\etal\ 1972) has been interpreted as precession by many authors (\eg,
Brecher 1972; Tr\"umper \etal\ 1986; \v{C}ade\v{z}, Gali\v{c}i\v{c} \&
Calvani, 1997; Shakura, Postnov \& Prokhorov 1998).

The fluid of the neutron star inner crust is expected to form a
neutron superfluid threaded by vortices. Above a fluid density of
$\sim 10^{13}$ g cm$^{-3}$, an attractive vortex-nucleus interaction
could pin vortices to the lattice. Vortex pinning in the
crust has been a key ingredient in models of the spin jumps, {\em
glitches}, observed in many radio pulsars (\eg, Lyne, Shemar \& Smith
2000).  In all candidate precessing neutron stars, the putative
precession is of long period, months to years, which is in dramatic
conflict with the notion of vortex pinning in the neutron star's inner
crust. As demonstrated by Shaham (1977), vortex pinning exerts a large
torque on the crust that causes the star to precess with a period
$(I_c/I_p)p$, where $I_p$ is the moment of inertia of the pinned
superfluid, $p$ is the spin period and $I_c$ is the moment of inertia
of the solid crust and any component coupled to it over a timescale
$\ll p$. If pinning occurs through most of the inner crust (as assumed
in many models of glitches), $I_p$ constitutes about 1\% of the entire
star. Though the coupling of the solid to the core is not well-known,
$I_c$ cannot exceed the star's total moment of inertia, so the 
precession period is {\em at most} $\sim 100$ spin periods. Hence, if
the long-period periodic behavior seen in the precession candidates
really represents precession, {\em the crustal vortices cannot be
pinned.} In this paper we present a resolution to this puzzle. We show
that vortices initially pinned to the inner crust would probably be
unpinned by the forces exerted on them by a crust set into
precession. In the absence of precession, vortices could still pin to
the inner crust with sufficient strength to account for giant pulsar
glitches.

The outline of the paper is as follows. In Section 2 we review the dynamics
of precession with pinning. In Section 3 we calculate the Magnus forces on
a pinned vortex in a precessing neutron star, and show that the Magnus
force per unit length is $\sim 10^{17}$ dyne cm$^{-1}$ throughout most
of the crust of PSR B1828-11, nearly two orders of magnitude larger
than the minimum force on pinned vortices in Vela just prior to a
giant glitch. Thus, if we assume that the force/length $f_p$ required
to unpin a superfluid vortex is in the range $10^{15}\lap f_p \lap
4\times 10^{16}$ dyne cm$^{-1}$, a consistent picture emerges for PSR
B1828-11 (and the other long-period precession candidates mentioned
above): {\em the precessional motion itself unpins the vortices and
keeps them unpinned}. In Section 4 we show that this interpretation makes
sense theoretically; we estimate the force/length required to unpin a
vortex in the crust, and find $f_p\sim 10^{16}$ dyne cm$^{-1}$. We
summarize our findings in Section 5.

\section{Free Precession with Pinning}

The problem of precession of the neutron star crust with a pinned
superfluid has been studied by Shaham (1977), Alpar \& Pines (1985),
Sedrakian, Wasserman \& Cordes (1999) and Jones \& Andersson
(2001). We revisit the problem here to emphasize key results for later
use. 

First, we briefly discuss the role that dissipative coupling between
the core liquid and solid would play in precession. Precession creates
time-dependent velocity differences between the crust and liquid that
vary over the star's spin period.  If the coupling time $\tau_{cc}$
between the crust and the core liquid is much longer than the crust's
spin period $p$, the precession will damp over $\simeq
2\pi\tau_{cc}/p$ precession periods (Sedrakian, Wasserman \& Cordes
1999). Coupling of the solid to the core liquid is not well
understood. Magnetic stresses allow angular momentum exchange between
the solid and the charged components of the core (Abney, Epstein \&
Olinto 1996; Mendell 1998), though this process is not by itself
dissipative. If the core magnetic field, of average strength $B$, is
confined to superconducting flux tubes, the crossing time for Alfv\'en
type waves through the core is $t_A\sim 4B_{12}^{-1/2}$ s for a
density of $10^{15}$ g cm$^{-3}$. The effective coupling time between
the crust and core charges cannot be less than $t_A$. The coupling
time between the neutron component of the core and the charges could
exceed $\simeq 400$ rotation periods (Alpar \& Sauls 1988).  Hence,
the timescale $\tau_{cc}$ for the {\em entire} core to achieve
corotation with the solid could exceed many rotation periods.  The
core might therefore be effectively {\em decoupled} from the solid as
the star precesses, with the crust precessing almost as if the core
were not there. Given the uncertainties, we will consider two regimes:
complete decoupling of the core liquid from the solid, and the
opposite regime of perfect coupling.

To study the precessional dynamics, we approximate the crust's
inertia tensor as the sum of a spherical piece, a centrifugal bulge
that follows the instantaneous angular velocity of the crust and a
deformation bulge aligned with the principal axis of the crust (Alpar
\& Pines 1985):
\begin{equation}
\bfI_c = I_{c,0} \bfdelta + \Delta I_\Omega \left (\nom\nom -
{1\over 3}\bfdelta\right ) 
+ \Delta I_d \left (\nd\nd- {1\over 3}\bfdelta\right ),
\end{equation}
where $I_{c,0}$ is the moment of inertia of the crust (plus any
components tightly coupled to it) when non-rotating and spherical,
$\bfdelta$ is the unit tensor, $\nom$ is a unit vector along the crust
angular velocity $\Om$, $\nd$ is a unit vector along the crust's
principal axis of inertia, $\Delta I_\Omega$ is the increase in
oblateness about $\Om$ due to rotation and $\Delta I_d$ is the {\em
deformation} contribution due to rigidity of the crust. We assume
$\Delta I_d\ll I_{c,0}$ and $\Delta I_\Omega\ll I_{c,0}$. For
simplicity, we take the crust superfluid to be perfectly pinned along
$\nd$. The total angular momentum of the crust plus pinned superfluid
is
\begin{equation}
\J = \bfI_c\cdot\Om + J_{sf}\nd,
\end{equation}
where $J_{sf}$ is the magnitude of angular momentum in the pinned
superfluid. For free precession $\J$ is conserved, with $\J$, $\Om$
and $\nd$ all spanning a plane (see Fig. 1). The {\em wobble angle}
$\theta$ between $\nd$ and $\J$ is a constant of the motion. The
precessional motion in the inertial frame can be seen by decomposing
$\Om$ as (see, \eg, Andersson \& Jones 2001)
\begin{equation}
\Om = \dot{\phi}\nJ + \dot{\psi}\nd,
\end{equation}
where $\nJ$ is a unit vector along $\J$, $\phi$ and $\psi$ are Euler
angles and overdots denote time differentiation. For small wobble
angle, the motion is given by $\dot{\phi}\simeq J/ I_{c,0}$ and
$\dot{\psi}\simeq-(\Delta I_d\Omega+J_{sf})/I_{c,0}$. The precession
is a superposition of two motions: 1) a fast wobble of $\nd$ about
$\nJ$, with a constant angle $\theta$ between the axes, and, 2) a
retrograde rotation about $\nd$. In the body frame, both $\J$ and
$\Om$ rotate about $\nd$ at frequency $\dot{\psi}$, the body-frame
precession frequency. For an emission beam axis fixed in the star and
inclined with respect to $\nd$, modulation at frequency $\dot{\psi}$
is observed.

For insignificant pinning ($J_{sf}\ll\Delta I_d\Omega$), the
body-frame precession frequency reduces to the classic elastic-body
result: $\dot{\psi}\simeq -(\Delta I_d/I_{c,0})\Omega$. However, for
significant pinning ($J_{sf}\gg\Delta I_d\Omega$), the precession is
much faster, $\dot{\psi}\simeq -J_{sf}/I_{c,0}$, as originally shown
by Shaham (1977). Taking $J_{sf}=I_p\Omega_s$, where $I_p$ is the
moment of inertia of the pinned superfluid and $\Omega_s$ is the
magnitude of its angular velocity,
$\dot{\psi}=-(I_p/I_{c,0})\Omega_s\simeq -(I_p/I_{c,0})\Omega$. If the
entire crust superfluid is pinned and $I_{c,0}$ is the moment of
inertia of the solid only, $I_p/I_{c,0}$ is $\simeq 2$ for most
equations of state, giving extremely fast precession with
$\dot{\psi}\simeq -2\Omega$. If pinning is imperfect, but vortices move
with respect to the lattice against a strong drag force, the
precessional dynamics resembles that for perfect pinning (Sedrakian,
Wasserman \& Cordes 1999).

The angle $\theta^\prime$ between $\Om$ and $\J$, is also a constant
of the motion (see Fig. 1). For $\theta$ and $\theta^\prime$ both
small, these two angles are related by
\begin{equation}
\theta^\prime\simeq {\left (\Delta I_d\Omega + J_{sf}\right )\over 
J  - J_{sf}- \Delta I_d\Omega}\theta.
\label{thetap}
\end{equation}
For insignificant pinning ($J_{sf}\ll\Delta I_d\Omega$),
$\theta^\prime$ is $\simeq (\Delta I_d/I_{c,0})\theta\ll\theta$.  For
significant pinning ($J_{sf}\gg\Delta I_d\Omega$),
$\theta^\prime\simeq(J/J_{sf}-1)^{-1}\theta$.  If the core is
decoupled from the solid, $J_{sf}$ can be $\simeq J$, and
$\theta^\prime$ can exceed $\theta$. If the entire inner crust
superfluid is pinned, $J_{sf}/J\simeq I_s/(I_s+I_i)$, where $I_s$ is
the moment of inertia of the crust superfluid and $I_i$ is the moment
of inertia of the lattice. For most equations of state, $J_{sf}/J$ is
about $0.7$ in this case, giving $\theta^\prime\simeq 2\theta$. On the
other hand, if the crust and core are tightly coupled,
$\theta^\prime\simeq (J_{sf}/J)\theta\lap 10^{-2}\theta$.

\section{Forces on the Vortex Lattice in a Precessing Star}

We have seen that significant vortex pinning produces a precession
frequency that is orders of magnitude faster than the periodicities
observed in PSR B1828-11 and other candidate precessing pulsars. We
now study the stability of the pinned state in a precessing star. When
making estimates, we take $f_p$ (the maximum pinning force/length that
the crustal nuclei can exert on the pinned superfluid vortices) to be
constant through the crust.

We begin by estimating $f_p$ from the angular momentum requirements of
giant glitches in pulsars.  The standard explanation for giant
glitches -- that they represent transfer of angular momentum from
the more rapidly rotating inner crust superfluid to the crust via
catastrophic unpinning (Anderson \& Itoh 1975) -- yields a lower limit
for $f_p$. As the stellar crust slows under electromagnetic torque,
vortex pinning fixes the angular velocity of the crust superfluid. As
the velocity difference grows, a Magnus force develops on the pinned
vortices. If the crust and superfluid are rotating about the same axis
(and therefore not precessing) the force per unit length of vortex is
$f_m=\rho_s\kappa r_\perp\omega$, where $\rho_s\simeq 10^{14}$ g
cm$^{-3}$ is the superfluid density, $\kappa\equiv h/2m_n$ is the
quantum of circulation ($m_n$ is the neutron mass), $r_\perp$ is the
distance of the vortex line from the rotation axis and $\omega$ is the
angular velocity lag between the crust and the pinned superfluid (see
eq. \ref{magnus} below). For a critical value $f_{m,c}=f_p$, vortices
cannot remain pinned. The corresponding lag velocity is
$\omega_c=f_p/\rho\kappa r_\perp$. Treating the crust as a thin shell
of constant density, the excess angular momentum stored in the pinned
superfluid (moment of inertia $I_p$) is $\Delta
J=(3\pi/8)(f_p/\kappa)(I_p/R)$.  Suppose that in a glitch some or all
of the excess angular momentum is delivered to the crust; the crust
suffers a spin-up of $\Delta\Omega \le \Delta J/I_c$, where $I_c$ is
the moment of inertia of the crust plus any part of the star tightly
coupled to it over timescales much shorter than the timespan of glitch
observations. The glitch magnitude is as much as
\begin{equation}
{\Delta\Omega\over\Omega}\le {3\pi\over 8} {f_p\over\rho_s\kappa
R\Omega}\left({I_p\over I_c}\right), 
\end{equation}
yielding the following lower limit for the force on pinned vortices
just prior to a giant glitch:
\begin{equation}
f_p \ge 10^{15} \left ({\Delta\Omega/\Omega\over 10^{-6}}\right )
\left ({\Omega\over 80\mbox{ rad s$^{-1}$}}\right ) \left
({I_p/I_c\over 10^{-2}}\right )^{-1}. 
\end{equation}
Here $\Omega\simeq 80$ rad s$^{-1}$ for the Vela pulsar and
$\Delta\Omega/\Omega\simeq 10^{-6}$ is typically observed. Analyses of
glitches in Vela and other pulsars show that $I_p/I_c\gap 10^{-2}$
(Link, Epstein \& Lattimer 1999); we have taken $I_p/I_c=10^{-2}$ as a
fiducial value. If glitches relax $f_m$ to nearly zero, the above
lower limit is an estimate for $f_p$. By comparison, the maximum
amount by which the Magnus force can increase between glitches is
$\Delta f_m=\rho_s\kappa r_\perp \vert\dot{\Omega}\vert t_g$, where
$\vert\dot{\Omega}\vert$ is the crust's spindown rate and $t_g$ is the
average time interval between glitches. For Vela, $\Delta f_m\simeq
10^{15}$ dyne cm$^{-1}$.

We now compare $f_p\sim 10^{15}$ dyne cm$^{-1}$, which is sufficient
to explain giant glitches, to the Magnus forces on vortices in a
precessing neutron star.  Define a Cartesian coordinate system
$(x,y,z)$ fixed in the crust and centered on the star, and let
$(\nx,\ny,\nz)$ be the corresponding basis vectors.  The superfluid
flow $\Oms$ past a pinned vortex segment creates a Magnus force per
unit length of vortex at location $\rbf=(x,y,z)$ of (see, \eg, Shaham
1977)
\begin{equation}
\fbf_m = \rho_s\kappa\nz\times\left
([\Oms-\Om]\times \rbf\right ).
\label{magnus}
\end{equation}
For simplicity, we assume vortex pinning along $\nz$ and take the
superfluid angular velocity to be $\Oms=\Omega\nz$. Consider an
instant at which $\Om$ and $\nz$ lie in the $y-z$ plane. 
The angular velocity of the crust is then 
$\Om\simeq\Omega(\alpha\ny+\nz)$, where
$\alpha\equiv\theta+\theta^\prime$.  The instantaneous Magnus force
per unit length of vortex as a function of position in the star is
\begin{equation}
\fbf_m = \ny\rho_s\kappa\Omega\alpha z.
\end{equation}
For $\Omega=16$ rad s$^{-1}$ (PSR B1828-11), the inferred
$\alpha$ of 3$^\circ$ and a density $\rho_s=10^{14}$ g cm$^{-3}$,
$\vert\fbf_m\vert$ exceeds $10^{17}$ dyne cm$^{-1}$ at $z=R$, a factor
of $\sim 100$ larger than the minimum force on vortices before a giant
glitch in Vela.  A vortex segment must unpin if $f_m>f_p$, or,
\begin{equation}
{\vert z\vert\over R} > {f_p\over\rho_s\kappa
R\Omega\alpha}\equiv {h\over R}, 
\label{unpin}
\end{equation}
where $R$ is the stellar radius.  For $\alpha=3^\circ$ and $\Omega=16$
rad s$^{-1}$ (PSR B1828-11), $f_p=10^{15}$ dyne cm$^{-1}$ and
$\rho_s=10^{14}$ g cm$^{-1}$, $h$ is $\simeq 0.007R$; only vortex
segments in a region of height $h\ll R$ are not unpinned directly by
the Magnus force (see Fig. 2). We next consider whether the vortex
array can be in static equilibrium with respect to the crust when only
a portion of the array is pinned very near the equatorial plane. We
find that the Magnus force on the unpinned segments exerts a torque
about $z=0$ which leads to further unpinning.

As vortex segments unpin at $\vert z\vert\ge h$, the angle between
$\nz$ and $\nom$ will assume a new value $\alpha^\prime<\alpha$. The
value of $\alpha^\prime$ will depend on how much pinning there is
initially, but it will not become less than $\theta$ (corresponding to
$J_{sf}$ becoming effectively zero; recall discussion following
eq. \ref{thetap}). The precession frequency will also be less than
before, as there is less pinned vorticity to exert torque on the
crust.  If the vortex segments are to remain anchored at $\vert z\vert
< h$, the vortex array must bend under the Magnus force exerted on
unpinned segments.  The extent to which vortices can bend is
determined by their self-energy or {\em tension}, which arises
primarily from the kinetic energy of the flow about them. For a single
vortex, the tension is $T\simeq(\rho_s\kappa^2/4\pi)\ln (r_v k)^{-1}$,
where $r_v$ is the vortex core dimension and $k$ is the characteristic
wavenumber of the bend in the vortex and $kr_v\ll 1$ (Sonin
1987). This tension, which is $\sim 10^8$ dyne, has a negligible
effect on the dynamics if the vortex is bending over macroscopic
dimensions. However, if a bundle of vortex lines bends, the effective
tension per vortex is enormous. A bundle of $N$ vortices has $N$ times
the circulation of a single vortex, and hence $N^2$ times the tension:
$T_N\simeq N^2 \rho_s\kappa^2/4\pi)\ln(r_v k)^{-1}$. The effective
tension per vortex is $T_{\rm eff}=T_N/N$. We estimate the angle
$\beta$ through which the vortex array can bend by taking the pinning
at $\vert z\vert < h$ to be infinitely strong, and seek a new static
configuration of a vortex line with tension $T_{\rm eff}$ under the
Magnus force. We will find that $\beta\ll\alpha^\prime$, so we
approximate the Magnus force in the new equilibrium as unchanged by
the small displacement of the line away from its original pinning
axis.  Let $\bfu(z)$ denote the displacement of a section of a vortex
from its original pinning position along $\nz$. In static equilibrium,
the shape of the vortex in the region $\vert z\vert\ge h$ is given by
\begin{equation}
T_{\rm eff} \bfu^{\prime\prime}(z) = \ny\rho_s\kappa\alpha^\prime\Omega z.
\label{equil}
\end{equation}
Let a vortex follow $\nz$ in the region $\vert z\vert <h$, so that
$\bfu^\prime(\pm h)=0$. Integrating eq. (\ref{equil}) once gives the
angle between a section of vortex and $\nz$ of
$\beta\simeq\rho_s\kappa\alpha^\prime\Omega (z^2-h^2)/2T_{\rm
eff}$. The number $N$ of vortices that must bend is $\simeq n_v2\pi
R\Delta R$, where $n_v=2\Omega_s/\kappa$ is the vortex areal density
and $\Delta R$ is the thickness of the pinning region (approximately
the thickness of the inner crust). The longest vortices that pass
through the region $x-y$ plane in the inner crust extend to a height
$z_0\simeq\sqrt{2R\Delta R}$. Evaluating $\beta$ at $z=z_0/2$ gives
$\beta\simeq \alpha^\prime/4\ln(r_v k)^{-1}$. Taking a bending
wavenumber $k=1/z_0$ gives $\beta\simeq 0.006\alpha^\prime$.  Hence,
the vortex array is far too stiff to bend over an angle
$\simeq\alpha^\prime$. This means that the unpinned segments are
prevented by tension from assuming a new static configuration in which
the Magnus force is small {\em unless} further unpinning occurs at
$\vert z\vert <h$. We now show that further unpinning is likely.

We treat individual vortices as infinitely stiff {\em when the entire
vortex array is bending}. The Magnus force on a vortex exerts a torque
about $z=0$. A segment can remain pinned and in static equilibrium
with respect to the crust only if pinning forces in the region $\vert
z\vert< h$ can exert a compensating torque. Suppose a given vortex
line has a length $H>h$. The Magnus force on this line exerts a torque
about $z=0$ of
\begin{equation}
\Nbf_m = \int_{-H}^H dz\, {\mathbf r}\times {\mathbf f_m} =
{2\over 3}\nx\rho_s\kappa\Omega\alpha^\prime H^3.
\end{equation}
If the section of the line at $-h\le z\le h$ is pinned, the lattice
can exert a compensating pinning torque of {\em at most}
\begin{equation}
\Nbf_p =2 \nx\int_0^h dz\, z f_p = \nx f_p h^2, 
\end{equation}
assuming the crust does not crack. Taking
$f_p=\rho_s\kappa\Omega\alpha h$ (eq. \ref{unpin}), the torque on the
unpinned segment exceeds that by the lattice if the length of
the segment satisfies 
\begin{equation}
H > \left ({3\alpha\over 2\alpha^\prime}\right )^{1/3} h. 
\label{H}
\end{equation}
These segments ``unzip'' from their pinning bonds, and are forced
through the lattice by the Magnus force. Segments shorter than $H$
exist only in the outermost region of the inner crust, in a region of
extent $\delta R\simeq H^2/2R$ in the $x-y$ plane (see
Fig. 2). Combining eqs. (\ref{unpin}) and (\ref{H}), we estimate the
extent of the pinning region to be
\begin{equation}
{\delta R\over R}\simeq {1\over 2} \left ({3\over
2\alpha^\prime\alpha^2}\right )^{2/3} \left ({f_p\over\rho_s\kappa
R\Omega}\right )^2. 
\end{equation}
The pinning region is largest if the core is decoupled, and if
unpinning changes $\theta^\prime$ from $\simeq 2\theta$ to
$\ll\theta$; in this case $\alpha^\prime\simeq\alpha/3$. For
$\alpha=3^\circ$ and $\Omega=16$ rad s$^{-1}$, we obtain $\delta R<1$
m. Pinning in this outermost region of the crust probably cannot occur
at all, but if it does, the moment of inertia of the pinned superfluid
in this small region would be too small to significantly affect the
spin dynamics of the star. If the core is tightly-coupled to the
crust, $\delta R$ is smaller by a factor of $\simeq 2$. Our estimate
for $f_p$ applies in the regions of the crust where pinning is
expected to be strongest; hence vortex unpinning is likely to be more
effective almost everywhere else in the crust. Vortex pinning is even
more difficult to sustain for precessing stars with higher spin
rates. 

The pinning strength $f_p$ can be considerably larger than $10^{15}$
dyne cm$^{-1}$ without affecting our conclusions. For example,
a pinning strength as large as $f_p=4\times 10^{16}$ dyne cm$^{-1}$
gives $\delta R=0.1R$ (assuming core decoupling), again probably
preventing significant pinning. We conclude that $f_p$ in the range
$10^{15}\lap f_p\lap 4\times 10^{16}$ dyne cm$^{-1}$ is sufficient to
explain giant glitches in young pulsars such as Vela, but insufficient
to sustain vortex pinning in PSR B1828-11. 

\section{Theoretical Estimate of the Vortex Pinning Strength}

The vortex-nucleus interaction arises from the density dependence of
the superfluid gap.  The details of this interaction are uncertain. In
the densest regions of the inner crust, where most of the liquid
moment of inertia resides, the interaction energy of a vortex segment
with a nucleus is estimated to be $E_p\simeq 5$ MeV (Alpar 1977;
Epstein \& Baym 1988; Pizzochero, Viverit \& Broglia 1997). The length
scale of the interaction is comparable to the pairing coherence length
$r_v\simeq 10$ fm, giving an interaction force of $F_p\simeq 5\times
10^6$ dyne per nucleus. Above a density $\simeq 10^{14}$ g cm$^{-3}$,
the vortex-nucleus interaction energy falls rapidly with density and
the vortex core dimension $r_v$ increases. The mass-averaged pinning
force is thus smaller than $5\times 10^6$ dyne; we will take
$F_p=10^6$ dyne as a fiducial value. Below a density of $\simeq
10^{13}$ g cm$^{-3}$, the vortex-nucleus interaction becomes
repulsive. Here vortices could pin to the interstices of the lattice,
but too weakly to play a significant role in the rotational dynamics
of neutron stars.

The degree to which vortices pin to the lattice nuclei is a complex
problem, but fortunately one where terrestrial analogs can provide
guidance. The pinning of elastic ``strings'' to attractive potentials
is a subject of current interest in condensed matter physics, arising,
e.g., in the pinning of magnetic vortices to lattice defects in type
II superconductors.  We follow the general reasoning of Blatter et
al. (1994) and D'Anna et al. (1997) to obtain a rough estimate of
$f_p$ for our problem.

Our estimate effectively treats the crust as an amorphous solid with
random pinning sites. This description of the solid is appropriate
according to recent calculations by Jones (1998b, 2001), though we
believe our pinning estimate to be roughly correct for a regular
lattice as long as the vortex does not closely follow one of
the lattice basis vectors.  In this context, it is important to
realize that pinning arises only because the vortex can bend. If the
tension $T$ were infinite, the forces on the vortex by nearby nuclei
would cancel on average (Jones 1991, 1998a).  On the other hand, if
the vortex tension $T$ were small compared to $F_p$, the vortex would
minimize its energy by adjusting its shape so as to intersect as many
pinning nuclei as possible. In this case the spacing between pinned
nuclei would equal the lattice spacing $b$ ($\simeq 30$ fm), and the
pinning force per unit length $f_p$ would be approximately
$F_p/b\simeq 3\times 10^{17}$ dyne cm$^{-1}$.  For superfluid vortices
in the NS crust, typically $F_p/T\simeq 10^{-2}$, so vortices bend
rather little, and $f_p$ is considerably below $F_p/b$,
as we now estimate.

The are essentially five physical parameters that together determine
$f_p$: The vortex tension $T$, the pinning force per nucleus $F_p$,
the vortex radius $r_v$, the nuclear radius $r_n$ and the typical
nuclear separation $b \equiv n_{nuc}^{-1/3}$. The tension of a single
vortex is
\begin{equation}
T = {\rho_s\kappa^2\over 4\pi}\ln(kr_v)^{-1},
\label{T}	
\end{equation}
where $k$ is the bending wavenumber.  For $r_n << r_v$, we expect the
value of $r_n$ to be unimportant. While $r_v$ and $r_n$ are comparable
in the denser regions of the crust (Pizzochero, Viverit \& Broglia
1997), in the rough estimate we make below we treat the nuclei as
points. We take as fiducial values: $F_p = 10^6$ dynes, $r_v=10$ fm,
and $b = 30$ fm.

If $T$ were infinite, so that the vortex could not bend toward nuclei,
the typical distance between nuclei along the vortex from random
overlaps would be $(\pi r^2_v n_{nuc})^{-1} = b (b^2/\pi r^2_v) \sim 3
b$.  For finite $T$, the vortex can bend to intersect extra nuclei,
but does so only on sufficiently long length scales, due to competition
between the attractive nuclear potentials and the elastic energy of
the deforming vortex.  Call these extra nuclei (that the vortex
intersects due to bending) the ``pinning nuclei''. The pinning nuclei
can bend the vortex over a {\em pinning correlation length} $L_p$. Let
$u$ be the transverse distance by which the vortex deviates from
straight over a length $L_p$.  For $u$ of order $r_v$ or greater,
the distance $L_p$ between successive pinning nuclei is of order $\sim
(n_{nuc}u^2)^{-1} = b (b/u)^2$.  The gain in binding energy due to
each pinning nucleus is of order $\sim F_p r_v$, while the energy cost
of bending is $\sim T (u^2/L_p^2) L_p$.  Bending to intersect an extra
nucleus becomes energetically favorable when these two energies are
comparable, giving a pinning correlation length of
\begin{equation}
L_p \sim b \left ({T\over F_p}\right )^{1/2} \left ({b\over r_v}\right
)^{1/2}.
\label{Lp}
\end{equation}
To calculate the vortex tension, we take the bending wavenumber to be
$k=1/L_p$. For a superfluid mass density of $\rho_s=10^{14}$ g
cm$^{-3}$ and the fiducial values given above, we solve eqs. [\ref{T}]
and [\ref{Lp}] simultaneously and find $L_p\simeq 20b$. The binding
energy/length of the bent vortex is $e_b \sim F_p r_v/L_b$. The
maximum Magnus force/length that the vortex can withstand before
unpinning is thus $f_p\simeq e_b/r_v \simeq F_p/L_p$, which for our
fiducial parameters is $f_p \sim 2\times 10^{16}$ dyne
cm$^{-1}$. The deviation of the bent vortex from straight is $u=(L_p
F_p r_v/T)^{1/2}$; this distance is comparable to $r_v$ for our
parameters, as we assumed {\em a priori}.

Glitch observations and the precession period of PSR B1828-11 suggest
a pinning strength in the range $10^{15}\lap f_p\lap 4\times 10^{16}$
dyne cm$^{-1}$. The above pinning estimate, though crude, shows that
such pinning strengths are theoretically sensible.

\section{Discussion}

In the most convincing example of pulsar free precession, PSR
B1828-11, we have shown that vortex pinning is unstable for a
reasonable pinning strength and wobble angle: the Magnus force on
pinned vortices is sufficient to unpin all of the vortices of the
inner crust. In support of this conclusion, we obtained in Section 4 a
theoretical estimate of the maximum pinning force/length $f_p$. The
large vortex tension increases the distance between effective
pinning sites relative to the case of small tension. We estimated
$f_p$ to be $\sim 10^{16}$ dyne cm$^{-1}$, smaller by a factor of
$\sim 10$ than the value obtained assuming a pinning spacing equal to
the lattice spacing. Our estimated $f_p$ is nevertheless large enough
to account for the giant glitches seen in radio pulsars, like
Vela. Our work here leads to at least one falsifiable prediction: {\em
PSR B1828-11, or any other pulsar, should not exhibit giant glitches
while precessing.}  (Small glitches could be explained by some
mechanism other than vortex unpinning, \eg, crustquakes).

In Section 3 we showed that partially-pinned vortex configurations cannot
be static. We did not attempt to solve for the dynamics of the
unpinned superfluid vortices or the effects on the precession of the
crust; we leave that as an interesting problem for future work.

\section*{Acknowledgments}

We thank R. I. Epstein, D. Chernoff and N.-C. Yeh
for helpful discussions. The work was supported in part by the
National Science Foundation under Grant No. PHY94-07194, NASA
grant NAG5-4093 and NSF MONTS Grant 436724.

\newpage

\begin{figure}
\caption{The constant angles in free precession. The symmetry axis
$\nd$ and spin axis $\Om$ span a plan containing $\J$. In the inertial
frame, $\nd$ and $\Om$ rotate about $\J$ at approximately the spin
frequency.}
\end{figure}

\begin{figure}
\caption{The Magnus force on a pinned vortex in a precessing
star. At heights $\vert z\vert>h$, the Magnus force exceeds the
pinning force per length of vortex, and the vortex segments are
unpinned directly by the Magnus force. Unpinned segments then torque
free most segments that extend through $z=0$, except for a small
annulus of lens-like cross section of height $H$ and thickness $\delta
R$ (shown with gray shading). For clarity, fewer vortices are shown on
the right side of the figure.}
\end{figure}


\begin{thebibliography}{}

\bibitem[a]{b}
Abney, M., Epstein, R. I. \& Olinto, A. V. 1996, ApJ, 466, L91. 

\bibitem[]{}
Alpar, M. A. 1977, ApJ, 213, 527. 

\bibitem[]{}
Alpar, M. A. \& Pines, D. 1985, Nature, 314, 334. 

\bibitem[]{}
Anderson, P. W. \& Itoh, N. 1975, Nature, 256, 25. 

\bibitem[q]{r}
Brecher, K. 1972, Nature, 239, 325. 

\bibitem[s]{t}
\v{C}ade\v{z}, A., Gali\v{c}i\v{c}, M. \& Calvani, M. 1997, A\&A, 324, 1005

\bibitem[ss]{ts}
\v{C}ade\v{z}, A., Vidrih, M., Gali\v{c}i\v{c}, M. \& Carrami\~{n}ana,
A. 2001, A\&A, 366, 930. 

\bibitem[u]{v}
Cordes, J. 1993, in {\sl Planets Around Pulsars}, ASP Conference
Series, Vol. 36, pp. 43-60 (Ed: Phillips, Thorsett \& Kulkarni). 

\bibitem[y]{z}
Desphande, A. A. \& McCulloch, P. M. 1996, in ASP Conference
Series, Vol. 105, p. 101 (Ed: Johnston, Walker \& Bailes). 

\bibitem[]{}
Epstein, R. I. \& Baym, G. 1988, ApJ, 328, 680. 

\bibitem[]{}
~---------. 1992, ApJ, 387, 276. 

\bibitem[]{}
Jones, P. B. 1991, ApJ, 373, 208 

\bibitem[]{}
~---------. 1992, MNRAS, 257, 501. 

\bibitem[]{}
~---------. 1998a, MNRAS, 296, 217

\bibitem[]{}
~---------. 1998b, MNRAS, 306, 327.

\bibitem[]{}
~---------. 2001, MNRAS, 321, 167. 

\bibitem[oo]{pp}
Jones, D. I. \& Andersson, N. 2001, MNRAS, 324, 811. 
 
\bibitem[]{}
Link, B. \& Epstein, R. I. 2001, ApJ, 556, 392. 

\bibitem[]{}
Link, B. \& Epstein, R. I. \& Lattimer, J. M. 1999, PRL, 83, 3362. 

\bibitem[]{}
Lyne, A. G., Shemar, S. L. \& Smith, F. G. 2000, MNRAS, 315, 534. 

\bibitem[ccc]{ddd}
Mendell, G. 1998, MNRAS, 296, 903. 

\bibitem[]{}
Pizzochero, P. M., Viverit, L, \& Broglia, R. A. 1997, PRL, 79,
3347

\bibitem[mmm]{nnn}
Sedrakian, A., Wasserman, I. \& Cordes, J. M. 1999, ApJ, 524, 341. 

\bibitem[ooo]{ppp}
Shabanova, T. V., Lyne, A. G. \& Urama, J. O. 2001, ApJ, 552, 321. 

\bibitem[]{}
Shaham, J. 1977, ApJ, 214, 251. 

\bibitem[sss]{ttt}
Shakura, N. I, Postnov, K. A. \& Prokhorov, M. E. 1998, A\&A, 331, L37. 

\bibitem[]{}
Sonin, E. B. 1987, Rev. Mod. Phys., 59, 87. 

\bibitem[yyy]{zzz}
Tannanbaum, H., Gursky, H., Kellog, E. M, Levinson, R., Schreier, E.,
Giacconi, R. 1972, ApJ, 174, L143. 

\bibitem[1]{2}
Tr\"umper, J., Kahabka, P, \"Ogelman, H., Pietsch, W., Voges, W. 1986,
ApJ, 300, L63. 


\end{thebibliography}
\end{document}